\documentclass[11pt]{article}
\usepackage{amssymb}

\newcommand{\half}{\mbox{$\frac{1}{2}$}}
\newcommand{\fslash}{\!\!\not\!}
\newcommand{\dslash}{\!\!\not\!\partial}
\newcommand{\tr}{\mbox{tr}}
\newcommand{\Tr}{\mbox{Tr}}

\begin{document}

\begin{table}
\begin{flushright}
IK--TUW 9912401
\end{flushright}
\end{table}

\title{Chiral anomalies and Poincar\'e invariance\thanks{Supported by
Fonds zur F\"{o}rderung der wissenschaftlichen Forschung,
P11387--PHY}}

\author{Jan B.\ Thomassen\thanks{E-mail: {\tt
thomasse@kph.tuwien.ac.at}} \\
{\em Institut f\"{u}r Kernphysik, Technische Universit\"{a}t Wien} \\
{\em A--1040 Vienna, Austria}}

\date{April 10, 2000}

\maketitle

\begin{abstract}

I study variations of the fermionic determinant for a nonabelian Dirac
fermion with external vector and axial vector sources. I consider
different regularizations, leading to different chiral anomalies when
the variations are chiral transformations. For these different
regularizations, I then consider variations associated with Poincar\'e
transformations. I find that both Lorentz and translational invariance
are anomalously violated in general, but that they are respected when
the variations of the determinant are regularized to give a
Wess-Zumino consistent anomaly (the Bardeen anomaly).  If the
variations are regularized to give a covariant anomaly, then
Poincar\'e invariance is not respected. Following Manohar in an
investigation of Poincar\'e anomalies in a chiral gauge theory, this
gives an alternative way to understand the need for a consistent
regularization of the variations of the fermionic determinant.

\vspace{\baselineskip}

\end{abstract}

\section{Introduction}

In many approaches to the problem of deriving low energy effective
theories of QCD we are required to calculate a variation of the
fermionic determinant with respect to a chiral transformation. For
example, in chiral perturbation theory we must construct the
Lagrangian for the pseudoscalar octet so as to transform in the same
way as the underlying QCD -- including the chiral anomalies
\cite{treiman}.

For the calculation of such a variation it is necessary to choose a
regularization, and in general different choices of regularization
leads to different expressions for the chiral anomalies. If we
consider for example a theory of a nonabelian Dirac fermion coupled to
external vector ($V$) and axial vector ($A$) sources, it is possible
to regularize the chiral variation of the fermionic determinant so
that the divergence of the vector current vanishes and the divergence
of the axial vector current is equal to the Wess-Zumino consistent
chiral anomaly \cite{wess}. Another possibility is to have covariant
anomalies \cite{bardeen-z}, in which case both the divergence of the
current and of the axial current are different from zero. (Note that I
am referring to anomalies for the divergence of the same axial current
for different regularizations, rather than anomalies for the
divergence of different currents within the same consistent theory,
which is perhaps more conventional.)

A consistent regularization ensures that the anomalies satisfy
Wess--Zumino consistency conditions. These consistency conditions are
integrability conditions for the variation of the determinant on gauge
orbits of chiral gauge transformations \cite{wess,ball}. When the
variation is regularized in another way, like the covariant, the
determinant itself it is not well defined. For example, it cannot be
obtained by integrating variations with respect to external fields.

In this paper I investigate the Poincar\'e transformation properties
of the nonabelian ``$V\!A$-theory'' mentioned above, and I will show
that for many other regularizations than the consistent one Poincar\'e
invariance is not respected. Rather, it is broken by anomalies, that
is, quantum effects. For example, the theory with covariant
regularization is not Poincar\'e invariant.

This provides an alternative way to understand the need for a
consistent regularization when variations of the fermionic determinant
is calculated. It may be more useful to have a symmetry argument for
this, rather than the perhaps more unfamiliar integrability condition.

An investigation slightly related to the one in this paper was carried
out some time ago by Manohar \cite{manohar} (see also \cite{nieh}). He
demonstrated that for a theory of a chiral fermion where the chiral
anomalies did not cancel, there were also Poincar\'e anomalies, and he
pointed out that this was another way to interpret the need for
anomaly cancellation.

It may also be worth pointing out that the Poincar\'e anomalies in
this paper are not directly related to the gravitational anomalies in
ref.\ \cite{alvarez-gaume}. First of all, the theory is regularized
consistently in that paper, and the fermionic determinant in is well
defined. Furthermore, the anomalies appear from Feynman diagrams with
a number of energy-momentum tensors at the vertices. The Poincar\'e
anomalies here, on the other hand, appear from diagrams with one
energy-momentum tensor and a number of vectors and axial vectors (for
the translational anomalies), or one angular momentum current and a
number of vectors and axial vectors (for the Lorentz anomalies).

The organization of the paper is the following. In sec.\ 2, I discuss
the nonabelian ``$V\!A$-theory'' on the classical level, and in
particular the symmetries of the model. In sec.\ 3, I discuss the
regularization of the variations of the fermionic determinant of the
$V\!A$-theory. The regularization scheme I use is proper time
regularization in Minkowski space, which is convenient here because
the exact specification of the regularization is controlled by a
``regularization operator'' $\tilde D$, related to the Dirac operator
$D$. I then discuss various choices for $\tilde D$. In sec.\ 4, I show
that the choice that leads to the Wess--Zumino consistent Bardeen
anomaly leads to Lorentz and translational invariance. On the other
hand, in sec.\ 5, I show that the choice that leads to the covariant
anomaly is not Poincar\'e invariant. Sec.\ 6 is a brief summary.

\section{Classical symmetries of the $V\!A$-theory}

The nonabelian $V\!A$-theory is chosen as a ``typical'' theory of a
Dirac fermion, and suitable for illustration of the main points. It is
given by the Lagrangian
\begin{eqnarray}
{\cal L} & = & \bar\psi(D+i\epsilon)\psi, \hspace{2em}
D \;=\; i\dslash-\fslash V-\fslash\! A\gamma_5-\mu.
\end{eqnarray}
Here, $V_\mu=V_\mu^at^a$ and $A_\mu=A_\mu^at^a$ are external vector
and axial vector sources, respectively, in the Lie algebra of some
group with generators $t^a$. I have added a small mass $\mu$ as an
infrared regulator, as well as an $i\epsilon$. I will usually suppress
both of these.

Nothing in our discussion of this model is not well known, until we
get to the part where we consider the Poincar\'e transformations in
the quantum theory. Note, however, that I have chosen to not treat the
external sources as gauge fields, since gauge invariance is not
important for our considerations.

The symmetries of this theory on the classical level include global
phase rotations and chiral rotations, translations and Lorentz
transformations, provided we assign the appropriate transformations
rules to the external fields $V_\mu$ and $A_\mu$. Thus we have phase
rotations
\begin{eqnarray}
\label{phase-rotation}
\nonumber
\psi & \to & e^{i\alpha}\psi, \\
\nonumber
\bar\psi & \to & \bar\psi e^{-i\alpha}, \\
V_\mu+\gamma_5A_\mu
  & \to & e^{i\alpha}(V_\mu+\gamma_5A_\mu)e^{-i\alpha},
\end{eqnarray}
with $\alpha=\alpha^at^a$, chiral rotations
\begin{eqnarray}
\label{chiral-rotation}
\nonumber
\psi & \to & e^{i\beta\gamma_5}\psi, \\
\nonumber
\bar\psi & \to & \bar\psi e^{i\beta\gamma_5}, \\
V_\mu+\gamma_5A_\mu
  & \to & e^{-i\beta\gamma_5}(V_\mu+\gamma_5A_\mu)
e^{-i\beta\gamma_5},
\end{eqnarray}
with $\beta=\beta^at^a$, translations
\begin{eqnarray}
\nonumber
\psi & \to & e^{ia_\mu P^\mu}\psi, \\
\nonumber
\bar\psi & \to & \bar\psi e^{-ia_\mu P^\mu}, \\
\nonumber
V_\mu & \to & e^{ia_\nu P^\nu}V_\mu e^{-ia_\nu P^\nu}, \\
A_\mu & \to & e^{ia_\nu P^\nu}A_\mu e^{-ia_\nu P^\nu},
\end{eqnarray}
where $P_\mu\equiv i\partial_\mu$ are the generators, and Lorentz
transformations
\begin{eqnarray}
\nonumber
\psi & \to & e^{i\frac{1}{2}\omega_{\mu\nu}J^{\mu\nu}}\psi, \\
\nonumber
\bar\psi & \to & \bar\psi
e^{-i\frac{1}{2}\omega_{\mu\nu}J^{\mu\nu}}, \\
\nonumber
V_\mu & \to & e^{i\frac{1}{2}\omega_{\mu\nu}J^{\mu\nu}}
V_\mu e^{-i\frac{1}{2}\omega_{\mu\nu}J^{\mu\nu}}, \\
A_\mu & \to & e^{i\frac{1}{2}\omega_{\mu\nu}J^{\mu\nu}}
A_\mu e^{-i\frac{1}{2}\omega_{\mu\nu}J^{\mu\nu}},
\end{eqnarray}
with generators $J_{\mu\nu} \equiv \half\sigma_{\mu\nu}+(x_\mu
i\partial_\nu - x_\nu i\partial_\mu) \equiv
S_{\mu\nu}+L_{\mu\nu}$. This list shows how the transformation rules
for the external sources are chosen. Note in particular that the phase
and chiral rotations are chosen for convenience to be global, since
gauge invariance is not relevant here.

Since these transformations are all symmetries, it means that the
Lagrangian transforms under an infinitesimal transformation by
\begin{eqnarray}
\delta{\cal L} & = & \partial_\mu\Lambda^\mu+\Delta
\end{eqnarray}
with $\Delta=0$. We shall see that in the quantum theory, for
regularizations that are not consistent, this is not automatically
true: Quantum effects lead to $\Delta\neq0$ for, e.g., the covariant
regularization -- also for the Poincar\'e transformations. Therefore,
in that case the Poincar\'e transformations are no longer symmetries
of the theory.

Finally, we may derive the classical conservation equations
\begin{eqnarray}
\nonumber
\partial_\mu J^{a\mu} & = & 0, \hspace{2em}
J_\mu^a \;\equiv\; \bar\psi\gamma_\mu t^a\psi, \\
\partial_\mu J^{a\mu}_5 & = & 0, \hspace{2em}
J_\mu^{5a} \;\equiv\; \bar\psi\gamma_\mu\gamma_5 t^a\psi,
\end{eqnarray}
where $J_\mu$ and $J_\mu^5$ are the fermion current and axial
current.

\section{The quantum theory}

The quantum $V\!A$-theory is given by the path integral
\begin{eqnarray}
Z & = & \int{\cal D}\psi{\cal D}\bar\psi e^{i\int d^4x
\bar\psi D\psi} \;\equiv\; \mathrm{Det} D,
\end{eqnarray}
which also formally defines the fermionic determinant. The effective
action $W$ ($Z\equiv e^{iW}$) is given by
\begin{eqnarray}
W & = & -i\Tr\ln D.
\end{eqnarray}
These expressions are formal and needs regularization to become well
defined. However, we are interested in the variations of the fermionic
determinant under some infinitesimal transformations. We should
therefore consider such variations and regularize these instead.

With infinitesimal parameters, the phase and chiral rotations of the
fermions induce a change in the Dirac operator
\begin{eqnarray}
\label{dirac-variation}
\nonumber
D & \to & e^{-i\alpha+i\beta\gamma_5}De^{i\alpha+i\beta\gamma_5} \\
\nonumber
  & = & D+i(D\alpha-\alpha D)+i(D\beta\gamma_5+\beta\gamma_5D) \\
  & \equiv & D+\delta D,
\end{eqnarray}
which in turn induces a change in $W$:
\begin{eqnarray}
\delta W & = & -i\Tr\delta D\frac{1}{D}.
\end{eqnarray}
The Jacobian $J$ is then determined by the requirement that the path
integral $Z$ is unchanged under a change of variables:\
\begin{eqnarray}
Z & = & Je^{iW+i\delta W} \;=\; e^{iW}.
\end{eqnarray}
Defining now a Lagrangian by $J \equiv \exp(i\int d^4x{\cal L}_J)$, we
have
\begin{eqnarray}
\int d^4x{\cal L}_J & = & -\delta W \;=\; i\Tr\delta D\frac{1}{D}.
\end{eqnarray}
${\cal L}_J$ thus contains the anomalies, $G_\alpha^a$ and
$G_\beta^a$,
\begin{eqnarray}
{\cal L}_J & \equiv & \alpha^aG^a_\alpha+\beta^aG^a_\beta,
\end{eqnarray}
and is therefore a quantity of interest.

When a regularization is specified, both the transformations
(\ref{phase-rotation}) and (\ref{chiral-rotation}) are in general seen
to be broken symmetries. We can then find the
conservation equations for the currents:
\begin{eqnarray}
\label{anomalies}
\nonumber
\partial_\mu J^{a\mu} & = & -G^a_\alpha, \\
\partial_\mu J^{a\mu}_5 & = & -G^a_\beta.
\end{eqnarray}
Since we know what the $G$'s look like for different regularization
schemes, at least the consistent and covariant ones -- see e.g.\
\cite{ball}, we will use eqs.\ (\ref{anomalies}) later (sec.\ 5 and 6)
to identify what scheme we are in.

Again note that, in the terminology used in this paper, the currents
are always defined in the same way in terms of the fermions, while the
anomalies varies with the regularization scheme. This is a difference
from the usual discussion of anomalies \cite{bardeen-z}, where the
regularization is assumed to be consistent while different definitions
of the currents are considered, each giving different expressions for
the anomalies.

Let us also record that for the Poincar\'e transformations, the
infinitesimal variations of the Dirac operator $D$ are
\begin{eqnarray}
\delta D & = & i(Da_\mu P^\mu-a_\mu P^\mu D)
\end{eqnarray}
for the translations, and
\begin{eqnarray}
\delta D & = & i(D\half\omega_{\mu\nu}J^{\mu\nu}
-\half\omega_{\mu\nu}J^{\mu\nu}D)
\end{eqnarray}
for the Lorentz transformations.

\section{Proper time regularization}

The regularization scheme I will use is proper time regularization in
Minkowski space \cite{thomassen-chiral}.  Traditionally proper time
regularization is used in connection with the Euclidean formalism, see
e.g.\ the review \cite{ball}. However, it turns out to be a great
advantage to work in Minkowski space, when different regularizations
-- consistent, covariant, etc.\ -- are discussed, since in that case
the various prescriptions are conveniently controlled by the choice of
a ``regularization operator'' $\tilde D$, related to the Dirac
operator $D$. This will be discussed in the next sections; see also
\cite{thomassen-chiral}. There is also no need to perform analytical
continuations of the fields and transformations in Minkowski
space. But other regularization schemes should of course be possible.

We can introduce a proper time integral and the operator $\tilde D$ in
the following way:
\begin{eqnarray}
\nonumber
\int d^4x{\cal L}_J & = & i\Tr\delta D\frac{1}{D} \\
\nonumber
  & = & i\Tr\delta D\tilde D\frac{1}{D\tilde D} \\
  & = & \int_{1/\Lambda^2}^\infty ds\Tr\delta D\tilde D
e^{is(D\tilde D+i\epsilon)}
\end{eqnarray}
The operator $\tilde D$ is a priori arbitrary, except that it must be
chosen to give the right $\epsilon$-prescription. This will then
ensure convergence at the upper integration limit. For the lower
integration limit the cutoff $\Lambda$ is introduced, which is to be
taken to infinity at the end of the calculation. When the appropriate
choice for $\tilde D$ is made, ${\cal L}_J$ will be regular and well
defined.

We can use the expression for $\delta D$ from eq.\
(\ref{dirac-variation}) to perform the proper time integral and write
$\int d^4x{\cal L}_J$ in a Fujikawa-like form
\cite{fujikawa,petersen}:
\begin{eqnarray}
\label{fujikawa-form}
\nonumber
\int d^4x{\cal L}_J & = & \int_{1/\Lambda^2}^\infty ds
\Tr\, i(D\alpha-\alpha D
+D\beta\gamma_5+\beta\gamma_5D)\tilde De^{isD\tilde D} \\
  & = & -\Tr\, \alpha\left(e^{i\tilde DD/\Lambda^2}
-e^{iD\tilde D/\Lambda^2}\right)
-\Tr\, \beta\gamma_5\left(e^{i\tilde DD/\Lambda^2}
+e^{iD\tilde D/\Lambda^2}\right).
\end{eqnarray}
Here I have used the cyclicity of the trace, the identity $\tilde
De^{isD\tilde D}D = \tilde DDe^{is\tilde DD}$, and the fact that only
the lower limit of the integral contributes due to the implicit
presence of the $\epsilon$.  Similarly, for the Lorentz
transformations and translations we have
\begin{eqnarray}
\label{fujikawa-poincare}
\nonumber
\int d^4x{\cal L}_J & = &
-\Tr\,\half\omega_{\mu\nu}J^{\mu\nu}
\left(e^{i\tilde DD/\Lambda^2}-e^{iD\tilde D/\Lambda^2}\right) \\
  & & -\Tr\,a_\mu P^\mu\left(e^{i\tilde DD/\Lambda^2}
-e^{iD\tilde D/\Lambda^2}\right).
\end{eqnarray}
To proceed from here it is necessary to choose a $\tilde D$.

\section{Consistent regularization}

Let us make the choice
\begin{eqnarray}
\nonumber
\tilde D & = & (i\gamma_5)D(i\gamma_5) \\
  & = & i\dslash-\fslash V-\fslash\! A\gamma_5+\mu.
\end{eqnarray}
Using the cyclicity of the trace and the fact that $\gamma_5$ commutes
with $\alpha$, we have
\begin{eqnarray}
\label{s-j-consistent}
\int d^4x{\cal L}_J & = & -2\Tr\,\beta\gamma_5e^{i\tilde
DD/\Lambda^2}.
\end{eqnarray}
The terms proportional to $\alpha$ in eq.\ (\ref{fujikawa-form}) have
thus canceled out.

We can find out which kind of chiral anomaly this choice of $\tilde D$
leads to.  After a standard calculation we get (see e.g.\ \cite{hu})
\begin{eqnarray}
\label{bardeen-anomaly}
\nonumber
{\cal L}_J & = & \frac{1}{4\pi^2}\tr\beta
\left[\epsilon_{\mu\nu\rho\sigma}
\left(\mbox{$\frac{1}{4}$}F_V^{\mu\nu}F_V^{\rho\sigma}
+\mbox{$\frac{1}{12}$}F_A^{\mu\nu}F_A^{\rho\sigma}\right.\right. \\
\nonumber
  & & \mbox{} \hspace{6em} -\mbox{$\frac{2}{3}$}i
A^\mu A^\nu F_V^{\rho\sigma}
-\mbox{$\frac{2}{3}$}iA^\mu F_V^{\nu\rho}A^\sigma
-\mbox{$\frac{2}{3}$}iF_V^{\mu\nu}A^\rho A^\sigma \\
\nonumber
  & & \mbox{} \hspace{6em} \left.-\mbox{$\frac{8}{3}$}
A^\mu A^\nu A^\rho A^\sigma\right) \\
\nonumber
  & & \mbox{} \hspace{4em} -\mbox{$\frac{2}{3}$}
\{D^V_\mu A_\nu,\{A^\mu,A^\nu\}\}
+\mbox{$\frac{1}{3}$}\{D^V_\mu A^\mu,A^2\}
+\mbox{$\frac{2}{3}$}i[D^V_\mu F_V^{\mu\nu},A_\nu] \\
  & & \mbox{} \hspace{4em} \left.-\mbox{$\frac{1}{6}$}
[F^A_{\mu\nu},F_V^{\mu\nu}]
+\mbox{$\frac{1}{3}$}D_V^2D^V_\mu A^\mu
- 2A_\mu D^V_\nu A^\nu A^\mu\right]
\end{eqnarray}
Here $F_V^{\mu\nu}$ and $F_A^{\mu\nu}$ are the Bardeen tensors,
\begin{eqnarray}
\nonumber
F^V_{\mu\nu} & \equiv & \partial_\mu V_\nu-\partial_\nu V_\mu
+i[V_\mu,V_\nu]+i[A_\mu,A_\nu], \\
F^A_{\mu\nu} & \equiv & \partial_\mu A_\nu-\partial_\nu A_\mu
+i[V_\mu,A_\nu]+i[A_\mu,V_\nu],
\end{eqnarray}
and $D^V_\mu\equiv\partial_\mu+i[V_\mu,\;\cdot\;]$. The terms
proportional to the $\epsilon$-tensor is the familiar Bardeen anomaly
\cite{bardeen}, while the other terms are of even intrinsic parity
\cite{balachandran}. The Bardeen anomaly is known to be consistent
\cite{wess}, hence the regularization resulting from this choice for
$\tilde D$ is a consistent regularization.

(Actually, this is not the whole story: Within the proper time scheme
one also gets a term proportional to $\Lambda^2$ (and terms
proportional to $1/\Lambda^2$, $1/\Lambda^4$,$\ldots$) in addition to
the $\Lambda$-independent terms in eq.\ (\ref{bardeen-anomaly}). It is
necessary to somehow remove this term if we intend to take $\Lambda$
to infinity at the end of the calculation. This can be done by the
Pauli--Villars inspired regularization described in \cite{hu}; see
also
\cite{thomassen-chiral}. This implies a ``theorem'' that we can simply
drop all $\Lambda$-dependent terms.)

It is easy to see that not only the terms proportional to $\alpha$ in
eq.\ (\ref{fujikawa-form}) cancel out, but also the terms proportional
to $a_\mu$ and $\omega_{\mu\nu}$ in eq.\
(\ref{fujikawa-poincare}). Hence the quantity $\Delta$ for the Lorentz
transformations and translations vanishes for this regularization, and
the theory is Poincar\'e invariant, as expected.

\section{Covariant regularization}

Let us now instead make the choice
\begin{eqnarray}
\nonumber
\tilde D & = & D^c \;\equiv\; -CD^TC^{-1} \\
  & = & i\dslash-\fslash V+\fslash\! A\gamma_5+\mu.
\end{eqnarray}
This is the ``charge conjugate'' of $D$; $C$ is the charge conjugation
matrix and transposition is with respect to the Dirac matrix
structure.

It is now no longer true that terms proportional to $\alpha$ in eq.\
(\ref{fujikawa-form}) automatically cancel. For a combination of a
phase rotation with parameter $\alpha=\alpha^at^a$ and chiral phase
rotation with parameter $\beta=\beta^at^a$ we get
\begin{eqnarray}
\nonumber
{\cal L}_J & = & \frac{1}{16\pi^2}\,\epsilon_{\mu\nu\rho\sigma}
\tr\left[\alpha(F_V^{\mu\nu}F_A^{\rho\sigma}
+F_A^{\mu\nu}F_V^{\rho\sigma})\right. \\
  & & \mbox{} \hspace{6em} \left.+\beta(F_V^{\mu\nu}F_V^{\rho\sigma}
+F_A^{\mu\nu}F_A^{\rho\sigma})\right].
\end{eqnarray}
This is the expression for the covariant anomaly \cite{ball}.

The presence of a term proportional to $\alpha$ means that the current
is not conserved. This is well known for a theory that is regularized
to have a covariant anomaly \cite{ball,treiman}.

This form of $\tilde D$ is similar to that of $D^\dagger$ in the
Euclidean formulation, where also the sign of $\;\fslash\!  A\gamma_5$
is changed relative to $i\dslash\;-\,\fslash V$. In the Euclidean case
the operator $D^\dagger$ has a special status, since it is used for
the construction of the positive operators $DD^\dagger$ and $D^\dagger
D$. The positivity of these operators will then ensure the convergence
of the upper limit of the proper time integral, instead of the
$i\epsilon$ which has the same effect in Minkowski space. For this
reason the Euclidean formalism produces ``naturally'' the covariant
anomaly, and it takes a considerable amount of work to produce other
anomalies, such as the consistent one \cite{ball}.

Within this regularization, we can now calculate the Jacobian that
corresponds to the translations and Lorentz transformations. The
procedure, which also includes some nonstandard elements, is
essentially the one described in ref.\ \cite{thomassen-chiral}, where
it was used in a slightly different context. The result is
\begin{eqnarray}
\label{l-omega}
\nonumber
{\cal L}_\omega & = & \frac{1}{8\pi^2}\omega_{\mu\nu}
\tr\left[\mbox{$\frac{1}{6}$}\square\tilde F^{\mu\nu}_A
+\mbox{$\frac{1}{6}$}i\partial_\rho[V^\rho,\tilde F^{\mu\nu}_A]
+\mbox{$\frac{1}{6}$}i\partial_\rho[A^\rho,
\tilde F^{\mu\nu}_V]\right. \\
  & & \mbox{} \hspace{5em} +\left.x^\mu(2V^\nu F_V\tilde F_A
+A^\nu F_V\tilde F_V)\right]
\end{eqnarray}
and
\begin{eqnarray}
\label{l-a}
{\cal L}_a & = & \frac{1}{8\pi^2}a_\mu\tr(2V^\mu F_V\tilde F_A
+A^\mu F_V\tilde F_V).
\end{eqnarray}
Even for $\omega_{\mu\nu}$ and $a_\mu$ constants, there are terms in
eqs.\ (\ref{l-omega}) and (\ref{l-a}) that are not total derivatives,
i.e.\ of the form $\partial_\mu\Lambda^\mu$. In other words,
$\Delta\neq 0$ for both Lorentz transformations and
translations. Thus, as advertized, Poincar\'e symmetry is violated in
this case.

\section{Conclusion}

We have seen that in the nonabelian $V\!A$-theory considered in this
paper, Poincar\'e symmetry survives quantization for a regularization
that leads to the consistent Bardeen form of the chiral anomaly, while
it is anomalously broken for a regularization that leads to the
covariant chiral anomaly. Of course many other choices for the
regularization operator $\tilde D$ can be made. (Indeed many other
choices than proper time regularization can be made for the scheme
itself.) In general Poincar\'e symmetry is violated since eq.\
(\ref{fujikawa-poincare}) does not vanish for general choices for
$\tilde D$.

Manohar, who investigated Poincar\'e anomalies in a chiral gauge
theory, pointed out that chiral anomalies and Poincar\'e anomalies
cancel simultaneously, since both are proportional to the
$d$-symbols. A similar result seems to be true here, i.e.\ that
$\Delta\neq 0$ for the Poincar\'e transformations whenever the
$\alpha$'s cancel out from eq.\ (\ref{fujikawa-form}). The
nonvanishing terms in $\Delta$ come from the derivative operators in
$P_\mu$ and $J_{\mu\nu}$.

\noindent
\paragraph{Acknowledgments} I thank Morten Krogh for reading and
commenting on a previous version of the manuscript.

\end{document}